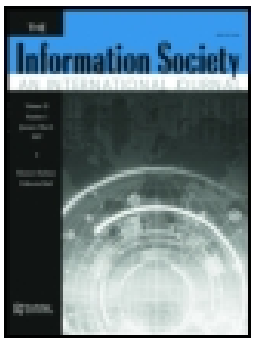

# Engineering Privacy by Design: Are engineers ready to live up to the challenge?

Kathrin Bednar, Sarah Spiekermann & Marc Langheinrich

Published online: 22 Mar 2019.

Submit your article to this journal ⏍

Article views: 47

View Crossmark data ⏍

# Engineering Privacy by Design: Are engineers ready to live up to the challenge?

Kathrin Bednar[a], Sarah Spiekermann[a], and Marc Langheinrich[b]

[a]Institute for Information Systems and Society, Vienna University of Economics and Business, Vienna, Austria; [b]Faculty of Informatics, Università della Svizzera Italiana, Lugano, Switzerland

## ABSTRACT

Organizations struggle to comply with legal requirements as well as customers' calls for better data protection. On the implementation level, incorporation of privacy protections in products and services depends on the commitment of the engineers who design them. We interviewed six senior engineers, who work for globally leading IT corporations and research institutions, to investigate their motivation and ability to comply with privacy regulations. Our findings point to a lack of perceived responsibility, control, autonomy, and frustrations with interactions with the legal world. While we increasingly call on engineers to go beyond functional requirements and be responsive to human values in our increasingly technological society, we may be facing the dilemma of asking engineers to live up to a challenge they are currently not ready to embrace.

Privacy is hardly a new topic. Over the years, a plethora of research and review articles as well as books on ethics and IT have pointed to the importance of privacy (Johnson 2009; Baase 2008; Vermaas et al. 2008; Culnan and Armstrong 1999; Acquisti, Brandimarte, and Löwenstein 2015; Bélanger and Crossler 2011; Smith, Dinev, and Xu 2011). There is also literature on how privacy can be undermined as well as protected by an appropriate system design (Friedman, Kahn, and Borning 2006; Cavoukian 2009; Spiekermann 2012; Spiekermann and Cranor 2009). In the policy arena, significant privacy regulation has been instituted since the 1980s (e.g. privacy guidelines of the Organisation for Economic Cooperation and Development 1980, and the Directive 95/46/EC of the European Parliament and Commission 1995). Furthermore, the 1990s saw a call for incorporation of privacy protection measures in IT products (e.g. Pretty Good Privacy, an encryption program for providing confidentiality for emails, developed by Phil Zimmermann and explained in Zimmermann 1995) and services (e.g. Privacy by Design, proposed by Hes and Borking 2000). Privacy by Design calls for proactive consideration and incorporation of privacy protection measures at the design stage of technological systems, setting privacy protection as the default option, and ensuring transparency of the collection, processing, transferring, and storage of personal data throughout the data lifecycle (Cavoukian 2010; Spiekermann 2012; Spiekermann and Cranor 2009). Therefore, Privacy by Design "requires the guts and ingenuity of engineers" (Spiekermann 2012, 39), as it is the systems engineers (i.e., software architects, information architects, interaction designers, product designers, and related specialities) who have to find a competent and creative way to realize privacy protection implementations. The central question of this article is: Are systems engineers ready to live up to this challenge?

More than twenty years ago, Smith (1994) investigated privacy management in the American corporate world. He found issues with all three societal mechanisms that typically influence corporate decisions. First, he found that the individual consumer is unable to exert pressure through the market, as consumers are often not informed about privacy intrusions, or it is not even clear to them what a privacy intrusion is. Second, the management lacks time and resources to proactively initiate corporate behaviours that protect privacy. And third, legislators lag behind technological developments with privacy regulations and they target privacy issues in a too narrow way, if they do so at all. For successful management of privacy in the future, Smith therefore called for a systemic fix, rather than a regulatory one.

What is the situation today? With the increasingly important role that the Internet and new information technologies play in our everyday lives, concerns about information privacy are growing. Consumer studies reveal that unease is spreading among citizens, as people fear losing control over their personal data. In the United States (Pew Research Center 2014) as well as in Europe (TNS Opinion & Social 2015), the majority of consumers feel that they have lost control over their personal data and are concerned about third party companies or the government accessing their personal information. At the same time, digital privacy breaches abound all over the world. Recent reports have revealed hundreds of data breaches in different sectors (e.g. banking, business, and health-care), which amounted to tens of millions of exposed records (Identity Theft Resource Center 2016; Verizon 2017). Regulators have started to react to these developments. In the U.S., new privacy regulations have been called for (The White House 2015) in addition to several sectorial privacy regulations (for a good overview, see Abramatic et al. 2015). In Europe, the new "General Data Protection Regulation" (GDPR; The European Parliament and the Council of the European Union 2016) enforces the protection of personal data. At the same time, personal data markets flourish more than ever before (Christl 2017) and personal data is considered the "new oil" of the digital economy (Schwab et al. 2011). Against this background, corporations find themselves torn between a rising call for more privacy-friendliness on one hand and the pressure to participate in the data economy on the other hand (Spiekermann et al. 2015). How does this situation influence the behaviours and attitudes of systems engineers (or "engineers" for short)? Have they become more aware of privacy issues? Have they assumed their responsibility and acquired the competences they need to build privacy-friendly systems? And are they provided within their corporations with the resources they need?

Very little is known about the subjective attitudes of systems engineers towards ethic-based practices such as Privacy by Design. Scholars have presented a holistic model of systems engineers' general job motivation (Sharp et al. 2009) and have looked at personality types of systems engineers (Cruz, da Silva, and Capretz 2015; Varona et al. 2012). But when it comes to the study of practical ethics-based design practices, the literature is sparse. Berenbach and Broy

(2009) have recently provided an analysis showing how organizational constraints impede engineers to behave in line with the code of ethics and professional conduct of the Association for Computing Machinery (ACM)[3]. In contrast, Szekely (2011), who studied a broader group of IT professionals, found that they live up to ethical demands if they are asked to do so by their organizations. They normally comply with decisions taken by their employers, regardless of whether these are in line with ethical conduct or not. However, none of these studies focus on privacy specifically.

Fifteen years ago, Langheinrich and Lahlou (2003) studied engineers' privacy behaviour to gather best-practice methods for incorporation of privacy protections in system design. They found that systems engineers rarely saw themselves as responsible for privacy protection measures. For the interviewees, privacy was "not yet necessary" as they first wanted to build prototypes. At the same time, privacy often turned out to be "no problem for prototypes". They saw privacy as "too abstract of a problem" that was "not necessary anymore" as security mechanisms like firewalls could take care of it. Langheinrich and Lahlou (2003) also reported that the engineers were "not feeling morally responsible" – they felt it was "not up to them" for a number of reasons, e.g. they lacked expertise. In some cases the interviewees said that privacy issues were simply "not part of deliverables" and correspondingly they did not have necessary time because it had not been allocated by their organizations. What is more, Birnhack, Toch, and Hadar (2014) point out that standard textbooks used in computer science education (e.g. Sommerville 2011) do not offer engineering students any timely knowledge on Privacy by Design. Instead, they reinforce the idea of maximizing data collection and minimizing the engineering effort on non-functional requirements.

More recent research seems to indicate that systems engineers' concern for the privacy protection has grown over the past few years. For example, computational modellers have stressed the importance of being faithful to reality and to users' values, as expressed in this statement of one modeller: "If we're going to produce models, they need to be accurate and they need to be useful. I don't want to lead people along the wrong path … They need to be grounded in a code of ethics. I think it's essential" (Fleischmann, Wallace, and Grimes 2010, 3). Similarly, Greene and Shilton (2018) found that an "ethic of care" for users is common among app developers. They concluded that developer forums such as the iPhoneDevSDK forum and the Android XDA forums act as quasi-regulators, setting privacy expectations for applications to be published on their platform stores and thereby guiding app developers' privacy efforts. A complementary study showed that certain work practices, such as navigating the platform's approval or user requests, can act as levers for privacy discourse, triggering larger debates on privacy and ethical requirements in general (Shilton and Greene 2017).

Yet, we have hardly any understanding of systems engineers' subjective attitudes toward ethical system design. We know little about their privacy related attitudes, beliefs, knowledge, skills and the degree of autonomy they have in organizations when it comes to the implementation of privacy protection measures. This gap in research calls for a comprehensive study of systems engineers' privacy related attitudes and engineering practices, which we are presenting in this article.

We conducted two complementary studies to investigate systems engineers' privacy related attitudes and engineering practices. First we conducted an in-depth qualitative study. Here we conducted 7.5 hours of semi-structured interviews with a small sample of senior systems engineers working for some of today's largest global software companies and renowned research institutions, the results of which will be presented hereafter. These interviews were complemented by a larger-scale survey-based study with 124 systems engineers (see Spiekermann, Korunovska, and Langheinrich 2018 for a full report on this study's findings). Both our qualitative and quantitative studies were guided by the Theory of Planned Behaviour (TPB; Ajzen 1985, 1991, 2002) as well as Jonas's work on the imperative of responsibility (Jonas 1984). This article focuses on the insights we gathered from our interviews, which provide a deep and nuanced understanding of the systems engineers' views on privacy from the engineering standpoint. We also report selected results obtained from the survey study by Spiekermann, Korunovska, and Langheinrich (2018) when they underscore insights from our interviews.

We adopted a mixed methods approach for the analysis of the interview data. We first applied a qualitative content analysis to inductively construct a system of categories and subsequently assessed how often a category was found in qualitative data from the interviews, thereby gaining a quantitative representation for each of the categories. We used the TPB as theoretical framework to understand systems engineers' ethical thinking within their organizational settings. Two back-to-back review articles that cover the empirical ethical decision-making literature from 1996 to 2011 have pointed out that the relationship between moral intent and moral behaviour has not been sufficiently studied and needs further empirical exploration (O'Fallon and Butterfield 2005; Craft 2013). As Ajzen's TPB predicts the link between intention and action, it is an appropriate theoretical framework for studying systems engineers' ethical decision-making. Other theories, such as the organizational legitimacy theory (Suchman 1995), also describe the relationship between an organization and its stakeholders. However, while organizational legitimacy theory focuses solely on attitudes, Ajzen's TPB models how attitudes are translated into behaviours.

In what follows, we first review the literature on engineers' privacy attitudes, beliefs, and work contexts as well as work autonomy. We then present the results from our interviews with four senior systems engineers and two heads of academic software groups. Our literature review and empirical results offer a deep insight into our interviewees' attitudes, emotions, and beliefs as well as their latitude regarding ethical decision-making within their organizational context.

## Relevant literature

The TPB states that the intention to engage in a specific behaviour is generally caused by three core factors: (1) people's instrumental and experiential attitudes towards a behaviour, (2) people's subjective norms, and (3) their perceived behavioural control. For our study context, this translates into engineers' intention to engage in Privacy by Design as a result of their attitudes towards information privacy, their personal and professional environment, and their degree of control over their systems' design. For the purpose of this interview study (as well as the consecutive survey study, see Spiekermann, Korunovska, and Langheinrich 2018), we defined privacy engineering as any activity undertaken by an engineer (i) to reduce the collection and storage of *personal* data (e.g. through data minimization or anonymization), (ii) to limit the sharing of personal data with third parties not explicitly authorized by the data subject, (iii) to give users full information about what happens to their personal data (i.e., transparency), and (iv) to give users real choice whether they consent to the processing of their personal data or not. We used the TPB to systematically review the literature on ethical engineering and structure our findings accordingly.

## Attitudes and beliefs

Attitudes towards a behaviour are experienced in two forms: instrumental attitudes determine if we find a behaviour useful and sensible; experiential attitudes determine if we find a behaviour enjoyable and pleasant (Ajzen 2006). Both forms of attitudes are typically driven by beliefs (Ajzen 1991).

The call for information privacy is met with scepticism and pessimism. In an age where a lot of their personal data is shared on the Internet, some people believe that "privacy is dead" (Heller 2011). Also, privacy is regarded as a value that needs to be traded off for more (national) security (Pavone and Delgi Esposti 2012; Bowyer 2004), transparency (Cochrane 2000; Mayes 2010) or knowledge (Land, Nolas, and Amjad 2004). Studies have found that privacy-friendly system designs can undermine functionality as well as convenience of a system for users (Nakayama, Chen, and Taylor 2016) as well as service administrators (Ciocchetti 2007). Following a Privacy by Design approach for a system is time-consuming and expensive, and does not support business goals that rely on accessing personal data (Krumay and Oetzel 2011). Furthermore, considering values in the modelling process can create conflicts between the goals and needs of the user, the client and the organization, between systems engineers' honesty and their obedience, as well as between (fast) product innovation and publication and the product's reliability and completeness (Fleischmann and Wallace 2010).

Privacy advocates are countering these negative observations by arguing that Privacy by Design can create business advantages (Hoffman 2014), reduce corporate liability (Ponemon Institute LLC 2011) and risks (Acquisti, Friedman, and Telang 2006) and does not necessarily undermine system security (Camenisch et al. 2005; Cavoukian 2009). They argue that privacy is a "fundamental right" (Solove 2008; Rouvroy and Poullet 2009), which is essential for functioning of democracies (Rouvroy and Poullet 2009) and trustworthy online environments in the future (Clarke 2001). Regulators have tended to follow this latter view, e.g. overhauling of the OECD Privacy Guidelines (Organisation for Economic Cooperation and Development 2013), passing of the General Data Protection Regulation law in Europe (The European Parliament and the Council of the European Union 2016), and efforts to build political privacy bridges, especially between the US and Europe (Abramatic et al. 2015).

All in all, ambiguous privacy beliefs and attitudes revolve around the value of privacy itself, its business impact, its technical practicability, its legal feasibility in a globalized IT world, and its potential conflict with other values. Even though the insights into engineers' individual thoughts are sparse as noted earlier, we must presume that – as part of a wider population – they are in the midst of this contradictory spectrum of views.

## Professional environment and subjective norm

Regardless of attitudes and beliefs, engineers are not as autonomous in their decisions regarding system design as they would like to be (Wallenstein 1974). The majority of systems are built in teams today, which can sometimes comprise more than 50 people. Therefore, the norms of behaviour reigning in such teams and the importance of team norms for the individual systems engineer could play a role in his or her propensity to consider privacy aspects. "Unless we look at and understand the social and institutional environment in which programmers work, attempts to hold the programmer solely accountable will be misguided", asserts Schaefer (2006, 1).

Ajzen (1985) referred to a social environment's influence on individuals as the *subjective norm*. He showed that the subjective norm (e.g. engineers' perceptions of what others expect of them), is a direct consequence of normative beliefs as well as an individual's motivation to comply with the norms and expectations that are common in the social environment. In our study context, this translates to whether or not the systems engineers believe that their employers and peers expect them to implement privacy requirements in their systems. These beliefs are weighted by the engineers' individual motivations to comply with these perceived norms and expectations.

Studies have provided support for IT professionals complying with the (ethical) requirements of their organizations. Shaw (2003) showed how IT professionals seek social consensus with their co-workers when it comes to difficult decisions regarding privacy. They "do not make ethical decisions in a vacuum, but instead look to their co-workers for guidance "and also consider the organizational effects of privacy engineering (such as additional cost expenditure) in their moral attitude towards privacy (Shaw, 2003). Szekely (2011) interviewed twelve IT professionals on their privacy engineering behaviour and surveyed 1,076 professionals in Hungary and the Netherlands. His findings regarding decision-making within organizations reveal that most engineers agree with privacy decisions made within a project and that they would "let it be known" if they disagreed; however, most of them stated that they would still implement decisions, even though they did not agree with them (Szekely 2011, 211).

So what kind of normative beliefs dominate in organizations? Do they encourage and/or enforce privacy-sensitive design? It seems reasonable to expect that today's organizations are cognizant of privacy as a design value. However, many organizations are operating in a highly competitive environment, which often pressures managers to support hype-driven technical innovation strategies (Spiekermann 2016). Berenbach and Broy (2009) discuss how this can lead to engineers having not enough time to deliver a software product, or having to deliver an incomplete product or a product with compromised quality; these and other dilemmas that engineers encounter at the work place do not nourish a working atmosphere where ethical considerations such as privacy concerns are being discussed.

### Perceived behavioural control

Perceived behavioural control deals with the "perceived ease or difficulty of performing the behaviour" (Ajzen 2002, 671). In our context, perceived behavioural control relates to the extent to which systems engineers feel that they have the freedom and capability to embed privacy mechanisms into a system. Control is determined by the form of IT governance in an organization (Webb, Pollard, and Ridley 2006). In official governance structures, managers often learn how valuable the craftsmanship and expertise of engineers are and that they should have the authority to find their own solution to a problem as they are the ones "closest to the work" Schaefer (2006, 3). Organizational factors determine how long a development effort is allowed to last. As a result of time and budget constraints, "the institutional workplace operates under the pressure of efficiency" (Schaefer 2006, 2). When engineering teams are put under pressure to deliver some software, they often do not have the time necessary to follow up on ethical requirements (Berenbach and Broy 2009). In a more recent study, Balebako et al. (2014) investigated privacy and security decision-making by app developers and found that smaller companies, which are constrained in time and resources, engage less in activities that promote information privacy and security, while larger companies advocate privacy or legal experts.

### Responsibility

Were ordinary people on the streets to be asked who is responsible for the design of IT systems, they would probably point their fingers to the engineers: "Engineers can influence the possible risks and benefits more directly than anybody else," notes Roeser (2012, 105). As long as human societies have engaged in tool-making and construction, there has been a recognition of the responsibility of the toolmaker for his creations. But this responsibility is not unambiguously accepted by engineers. Already in 1974, Wallenstein wrote in IEEE Spectrum: "We engineers may not appreciate being likened to slaves and prisoners, but where is our spirit of free men? Are not most of us slaves to job opportunities and pay checks, and prisoners of a system in which responsibilities are shouldered by others?" (Wallenstein 1974, 78). In 2006, Schaefer asked the question "Should the programmer be the one solely held accountable for the software faults?" (Schaefer 2006, 1). Similarly, Langheinrich and Lahlou (2003) reported that engineers do not feel morally responsible and that they felt it was not "up to them". Szekely (2011) found that IT professionals ultimately see the responsibility with the users, who are supposed to protect their personal data by using privacy-enhancing (protection) tools. He also found that "the majority of the respondents think that they bear no responsibility in ensuring the legality of the system they help to develop or run: the responsibility lies with either the management or the clients, but in any case outside their competency" (Szekely 2011, 209). These findings are not in line with the imperative of responsibility that engineers have been called to live up to by philosophers such as Hans Jonas (1979), nor do they match the code of ethics of major professional engineering associations such as the ACM[3] or the Institute of Electrical and Electronics Engineers (IEEE)[4].

### Methodology

We conducted six extensive interviews, spending roughly 7.5 hours with four senior systems engineers and two heads of academic software groups. We estimate that the totality of our interviewees have amassed more than 60 years of experience working for global software houses like Google, IBM, Alcatel Lucent, and Microsoft or doing research for leading ubiquitous computing research labs. They were all in senior positions that are usually attained only after many years of hands-on software and engineering experience. One of the authors conducted and digitally recorded the interviews at a major IT conference (Ubicomp, which is a conference on new and avant-garde technologies) with the informed consent of the interviewees. The participants' names were fully anonymized. The interviews were conducted in English and German (three German interviews, three English interviews); German interviews were translated into English by the authors.

In addition, 124 engineers answered an online survey that measured the scale of attitudes, subjective norm perceptions and control aspects (see Spiekermann, Korunovska, and Langheinrich 2018 for a detailed report of the survey study). Participants were recruited through a mailing list from the same IT conference, ensuring reach to engineers who are developing new systems rather than maintaining corporate infrastructures for which privacy designs may have been decided long ago. It took them 38 minutes on average to answer, participating in a lottery for Apple products and receiving Amazon vouchers in return. 81% of the respondents were male and on average 36 years old. Thirty nine percent ($n = 39$) from German-speaking countries, 13% ($n = 16$) from the US, 10% ($n = 12$) from Italy. The rest were comprised of 29 different nationalities from across the world. In terms of work position and environment, 77% ($n = 96$) the professional engineers and 23% ($n = 28$) PhD students. Sixty two percent ($n = 73$) work in a research-related environment (i.e., university, corporate R&D or research institutes), 48% ($n = 46$) in product development for an IT company, two for NGOs, and three for governments. Twenty five percent ($n = 29$) indicated having a leadership position. In this article we primarily focus on the results obtained from the interviews. That said, our qualitative findings are largely in keeping with the findings of the survey study (Spiekermann, Korunovska, and Langheinrich 2018), and where the two diverge, we discuss these divergences.

### Interview guide

We operationalised privacy and security engineering with the definitions provided earlier. Sharing these with interviewees we asked them to think about concrete ethical design targets in the past when answering our questions.

As outlined above, we used the TPB and Jonas's imperative of responsibility as guides for our semi-structured interviews (see Appendix 1 for the full interview guide). The interview guideline first focused on ethical decision-making in system design and development in general ("What is 'ethical computing' from your perspective?") and then focused on privacy and related security mechanisms in particular (e.g. "What are disadvantages and challenges of incorporating privacy mechanisms into your projects?"). It also included questions about our interviewees' (experiential) attitudes (e.g. "Do you find security problem solving more pleasing and enjoyable than privacy problems?"), their perceived social pressure or subjective norms (e.g. "What do most people who are important to you think about privacy and security?" and "How much do you want to comply with what your environment thinks?") as well as their perceived behavioural control (e.g. "Do you have the skill set?" and "Do you have the time?"). Inspired by works of Jonas (1979), we decided to also cover responsibility as an interview topic (e.g. "How do you see your own responsibility?").

### Analysis of interview data

Transcripts of the six interviews totalled 63 pages, comprising 34,290 words. We analysed the transcribed text passages in two phases using NVivo software (version 11), starting with an explorative and inductive content analysis. Based on the results of this first analysis phase, we deployed a descriptive and deductive analysis method (Mayring 2014).

In the explorative analysis phase, we marked 588 passages in the interview transcriptions (containing single words, phrases or sentences) as relevant. We then inductively generated themes from these text passages by identifying similarities and regularities. This first step yielded 14 *themes*. Ten of these 14 themes had less than 30 corresponding passages each. In contrast, the theme "privacy" had 243 corresponding passages, spanning almost half of the comments and statements (41%). In order to explore these data in a focused way, we focused solely on privacy in the second and main phase of the analysis (presented below).

In the second phase, we chose a descriptive approach for the content analysis (Mayring 2014). We categorized the pool of coded comments and statements that targeted privacy deductively, using the TPB framework as a guide, and registered how often each theme appeared in the interviews. Through this process, we developed six categories that correspond to TPB factors (privacy beliefs, instrumental attitudes, experiential attitudes, subjective norm, control beliefs, perceived behavioural control), and an additional category relating to responsibility (see Figure 1).

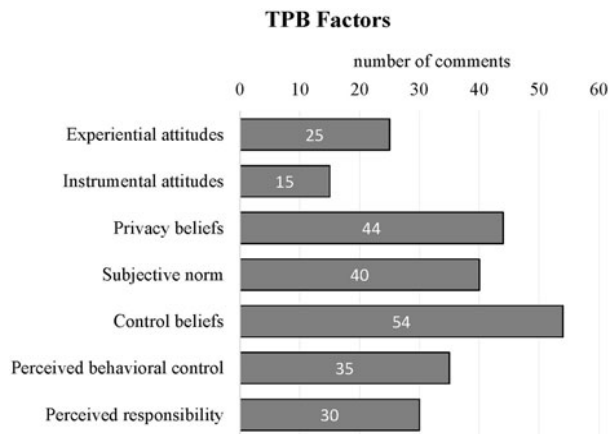

**TPB Factors**

While some of the interview questions targeted TPB factors directly – for example, the question "How do you spontaneously feel about ethical requirements?" corresponded to experiential attitudes – our interviewees did not always answer in a straightforward way. Often, our interviewees covered several TPB factors in one answer. Moreover, many statements that corresponded with a specific TPB factor did not come up with the corresponding question but at other points in the interview. Therefore, we always took the whole interview as a basis for the analysis, as opposed to focusing only on the questions corresponding with one TPB factor. We then systematically placed the statements in specific categories, drawing on the definition of each of the factors of the TPB outlined earlier.

It is important to understand that *privacy beliefs* differ from other TPB factors in that they manifest as general statements rather than expressions of subjective experiences – generic beliefs about the nature of (information) privacy or related concepts such as "consent". Therefore, wherever a statement was generic and did not express the interviewee's personal attitude or perception, we categorized it as a belief – either as a general privacy belief or as a more specific control belief their individual control over privacy implementation as engineers. We categorized all those comments and statements that focused on the importance of (information) privacy as *instrumental attitudes*. Whenever emotional adjectives were used by our interviewees, we categorized those comments and statements as *experiential attitudes* with a range from positive to negative. The *subjective norm* category – representing the perceived social pressure to behave in a certain way – encompasses all comments and statements that describe how engineers perceive the importance of information privacy in their working environment as well as in the general population. Statements referring to the engineers' own resources, time, knowledge, experience, capabilities or autonomy to solve privacy issues and implement privacy mechanisms were placed in the *perceived behavioural control* category. On the other hand, general statements about privacy, related concepts and aspects that have an influence on whether one perceives it as possible to protect information privacy by means of system design, were placed in the *control beliefs* category. All comments in which the engineers directly referred to their own or others' responsibility or tasks that they (or others) need to fulfil, as well as rules they need to comply with, were placed in the *responsibility* category.

Inter-coder agreement was secured via constant communication between a primary coder and a second coder who acted as the supervisor. This second coder had access to and was familiar with the whole interview material and the definitions of TPB factors. The supervising coder checked and confirmed the analyses of the first coder and wherever discrepancies were found, the two coders discussed the selection and interpretation of the respective text segments. While this kind of inter-coder agreement is described by Mayring (2014, 114) as a "'lighter' test," it allows for complete agreement between the two coders on the final assignment of all text segments in a system of categories.

## Findings

A word frequency analysis of all the interviews showed that the ten key words that were most often mentioned in the interviews by the interviewer and the interviewees were, in descending order, "privacy," "people," "data," "system," "product," "security," "information," "design," "user," and "location". While the interviews were initially structured to focus on ethical decision-making in system design and development in general and privacy and security mechanisms in particular, the actual interviews ended up focusing heavily on privacy, security being less eagerly discussed.

While the small number of interviewees limits the generalizability of our findings, it also allows for an in-depth analysis of the different subjective attitudes. Instead of determining how dominant one belief or attitude is within a representative sample, our study focuses on the different possible configurations of beliefs and attitudes and on what we can learn from them. For example, we illustrate how often several – even seemingly contradictory – attitudes, beliefs, and perceptions are held by one single person. For this reason, our results do not only depict the number of statements that fall within each of the categories [indicated in squared brackets], but also indicate who made these statements, whereby our interviewees are anonymously represented by the letters A to F hereafter.

The interviewees expressed their attitudes towards information privacy and its consideration in system design in 40 comments, out of which 25 comments expressed their experiential attitudes and 15 comments revealed how they evaluate the importance of privacy (instrumental attitude). The most relevant questions in the interview guide on this score were "How do you spontaneously feel about ethical requirements?" (targeting experiential attitudes), and "What is your own thinking?" (targeting instrumental attitudes). The categories that emerged comprise group statements indicating positive, neutral, and negative experiential attitudes towards the implementation of information privacy as well as positive and negative instrumental attitudes regarding its importance.

### Experiential attitudes

The experiential attitudes tend to be rather negative, as Figure 2 shows. It depicts the experiential attitudes in the interviewees' statements as well as the number of statements that fall within each category. The letters in the bars show which interviewee is represented in each of the categories.

Four of the six interviewees considered the incorporation of information privacy mechanisms somewhat "inconvenient" (mentioned six times by one of the interviewees) or otherwise negative ("not pleasing or enjoyable or exciting", "not enthusiastic", "it just becomes a nightmare") [10 comments]. Furthermore, it is demanding or (intellectually) "challenging" [8 comments] – something that can be good or bad, according to one remark.

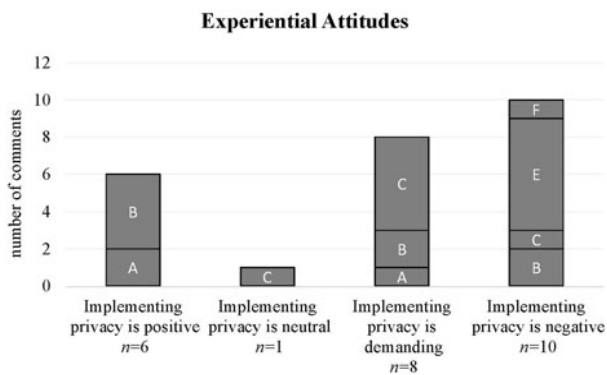

**Figure 2.** The engineers' experiential attitudes towards the incorporation of information privacy mechanisms as expressed in 25 comments; the interviewed engineers are anonymously represented within the bars as the letters A to F.

Two of the six interviewees found some positive words for the implementation of information privacy, saying that implementing privacy mechanisms makes them happy ("if it wants me to incorporate privacy I will be very much happy [sic]") or mentioning that it is "interesting", "exciting" and "satisfying" [6 comments]. However, as the letters in Figure 2 indicate, even these two interviewees (A and B) had mixed feelings towards privacy, as they equally mentioned negative aspects or expressed how demanding it is. One interviewee associated it with "neutral emotions" [1 comment].

This rather negative experiential attitude towards privacy was confirmed in our subsequent quantitative study (Spiekermann, Korunovska, and Langheinrich 2018), where a 5-point semantic differential scale with five bipolar adjective pairs was used to measure experiential attitudes (e.g. annoying - pleasing). The mean across adjective pairs was $M = 3.32$ $(SD = 0.82)$. Forty percent of the engineers surveyed do not like to engage in privacy engineering. Experiential attitude towards privacy engineering was significantly correlated with an engineer's belief that transparency would be more important as a value than privacy ($r = -.41$; $p < .001$), pointing to a value conflict. Those, however, who believed that privacy engineering was important to enable a power balance between corporations and citizens were also more likely to enjoy privacy engineering ($r = .22$; $p < .05$).

### Instrumental attitudes

When it comes to instrumental attitudes, the views are much more balanced (see Figure 3). Eight comments pointed to privacy being important and sensible while seven comments questioned its importance. It is important to note that three of the interviewees [C, E, and F] hold both views.

Five out of six interviewees pointed to the importance of information privacy [8 comments], saying that it is "sensible", "relevant" or "(very) important" that "design and human interaction issues are increasingly accepted as a critical aspect of any software that we develop" and that they are "very concerned" about it. However, three out of the five interviewees who mentioned the importance of information privacy at some other point of the interview, also made comments expressing that privacy is not important nowadays ("Now privacy is not as big as then", "and regarding up-to-date: this is more a general question, if we are going to see it a lot; and I believe while this will not be the all-determining topic in two, three years, it will yet be important") and referred to information privacy as "secondary" or "side part", e.g. when compared to Internet connection or functionality [7 comments]. Thus, some of our interviewees effectively contradicted themselves, giving the impression that the engineers are split in their views.

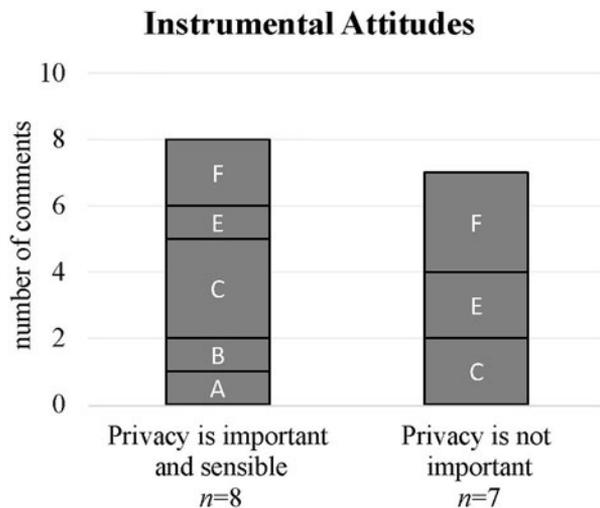

**Instrumental Attitudes**

**Figure 3.** Comments expressing the engineers' instrumental attitudes towards the incorporation of information privacy mechanisms, $n = 15$; the interviewed engineers are anonymously represented within the bars as the letters A to F.

Our survey results (Spiekermann, Korunovska, and Langheinrich 2018) underscore the interview results that instrumental attitudes are much more positive. In the survey, we used a 5-point semantic differential scale with six bipolar adjective pairs to measure instrumental attitudes (e.g. privacy engineering is worthless - valuable). The mean across adjective pairs was $M = 4.18$ ($SD = .76$) and hence much higher than with experiential attitudes. Only a small fraction of 10% of the engineers find privacy engineering useless. Again, the conflicting value of transparency ($r = -.36$, $p < 0.1$) and the belief in corporate-citizen power balance ($r = .28$; $p < .05$) influence the attitude held.

All in all, the results show that the engineers' experiential attitudes towards information privacy are rather unfavourable and that their instrumental attitude is ambivalent.

### Privacy beliefs

There were 44 statements and comments related to privacy beliefs. These statements were often made in relation to the question "What is 'ethical computing' from your perspective?". Figure 4 displays the nine beliefs that run through their statements. All of these beliefs are critical, sceptic, or negative.

Four out of six interviewees thought that privacy is not an absolute value [9 comments], as it has "room for interpretation" and a "human element." It is not equivocally perceived as a fundamental right ("I

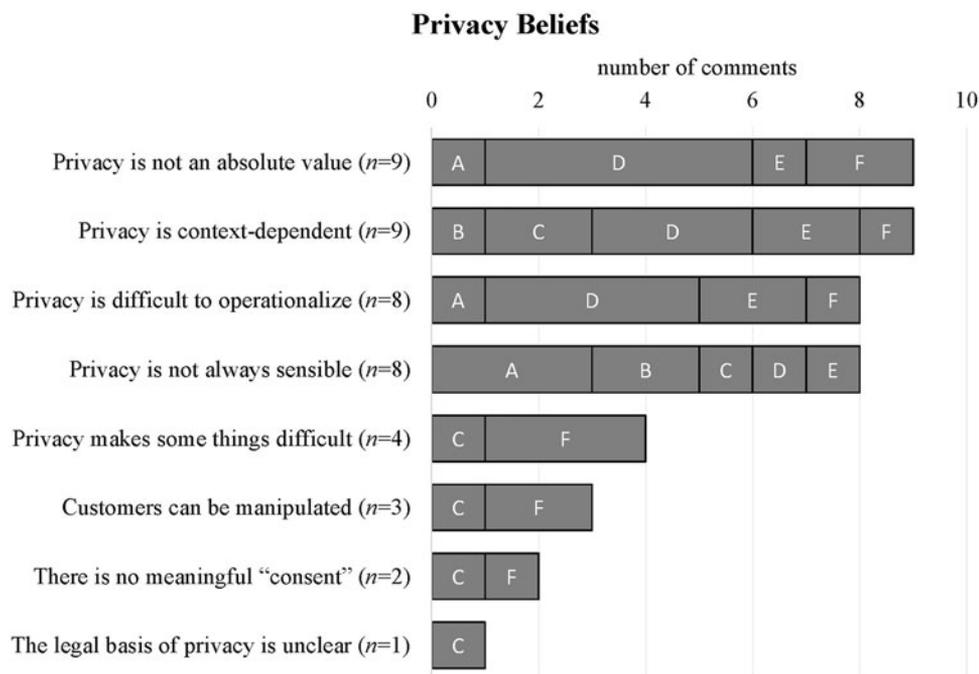

**Privacy Beliefs**

**Figure 4.** Engineers' beliefs regarding (information) privacy as expressed in 44 comments, ordered by the descending numbers of comments for each belief; the interviewed engineers are anonymously represented within the bars as the letters A to F.

cannot share the idea that privacy is a fundamental right that is just indefeasible") as it is merely "a perception to people [sic]" that always changes, a "commodity" that can – and in certain contexts *should* – be traded or sold ("if I own the data, I should sell the data").

Five out of six interviewees thought that privacy is context-dependent [9 comments]. They felt that in some contexts it is more important to consider privacy than in others; here they referred to companies making money with the data as opposed to the academic or research context ("And I think in the academic environment it is not as critical as in the company environment, where you make money with the data"; "From a research point of view there's nothing stopping us from doing something. I think this question becomes a lot more relevant when you are making a product"). They also believe that users assess their privacy differently in different contexts (" ….who can see it and who cannot. We did that with a study and there it was very clear, that you have to decide that as the case arises. Well, for example, 'is the user on the toilet or not' - this is a moment where I do not want to call"). The engineers' view on the legitimacy of information privacy also influences their ethical perception of their own actions ("I don't believe that collecting data per se violates privacy; there are many situations where we collect data"; "Well, it depends because if we are not misusing anything, if we are not selling this information to anybody … ").

The engineers do not always know how to operationalize privacy [8 comments]: "One privacy question here is: is it the collection of data the problem or the exposure of the data?", "If we approach systematically what we do, we lack understanding: what then is the overall system that we call privacy?". They expressed that privacy is "not as well formalized and understood" and that different engineers have different ideas and solutions.

Furthermore, engineers point to the issue that it is not always sensible to implement privacy [8 comments]. They mentioned that data is often needed for systems to work ("The system would need to collect data in order to do something meaningful"; "There are systems that only work when I have big data") as well as for other purposes like advertising ("on the other hand, you do want to use the mass of data for advertisement"). Another argument was that it not only protects individuals, but also gives citizens and customers the power of misuse ("maybe privacy is one thing, where the corporation is not misusing the data, but anonymity can let citizens misuse the corporation. What if I had anonymized phones, and I basically make a call and the corporation doesn't know who to bill?"; "Transparency can of course go in both directions, you cannot forget about that. And transparency can be the opposite of privacy. Full transparency also stands for more power on the customer's side").

In addition, two interviewees pointed out that privacy makes things difficult [4 comments] as it can slow down processes ("it could nevertheless be possible that decisions are delayed or processes slowed down at the code level"), impede functionality, and hinder research because less information is available ("you can have an access control list … that makes things very heavy, because in your data model you have to have meta data that describe your data").

Two interviewees mentioned issues related to users and customers. Firstly, they believe that the implementation of privacy becomes more tricky as customers can be manipulated and bribed by companies [3 comments]: "If I, as a customer, agree with the collection of my data, I cannot do anything against it; that means I can be bribed", "if we look into different other [sic] systems, if you have a very bad user interface but a very good functional system, it will still not work", "that [a bad user interface for privacy settings] is intentionally done to make people just ignore it".

Secondly, engineers are aware of the difficulty of giving and receiving meaningful consent [2 comments], pointing out that "the biggest lie we do every day [sic] is when we click the 'I agree' button; you never read those privacy statements and agreements" and "You can do these consent-things, but then the question arises if that is enough. Do the people really read and understand that we collect these data and analyse it for research?".

And lastly, one interviewee saw the legal basis of (information) privacy as unclear [1 comment]: " … this is not at all acknowledged by the data protection law; there are also very few court decisions that said: 'in this case, there was enough anonymization and in that case, there was not.'"

In the subsequent quantitative study (Spiekermann, Korunovska, and Langheinrich 2018), we tested for a larger number of beliefs, which we mostly identified from the literature. The full set of beliefs investigated as well as their correlations with privacy attitudes can be found in Appendix 2. However, the beliefs in the value of transparency and in the necessity to balance the power of corporations with that of citizens were confirmed as highly relevant in the quantitative survey study.

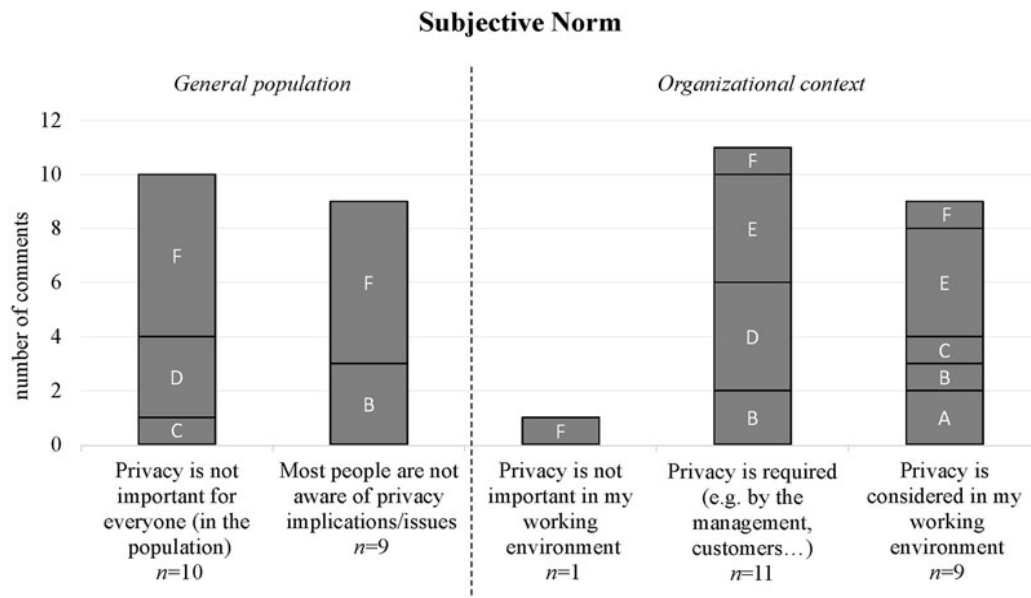

**Figure 5.** The engineers' perceived social pressure from the general population and their organizational context to incorporate information privacy mechanisms as expressed in 40 comments; the interviewed engineers are anonymously represented within the bars as the letters A to F.

### Professional environment and subjective norm

Figure 5 displays the engineers' subjective norms that emerged from 40 comments on perceived social pressure from their working environment and the general population. In relevant questions from the interview guide engineers were asked about their assumptions as to what their respective organization and people who are important to them think and expect as well as their own motivation to comply with these norms. Results show that engineers do not perceive any pressure from the general population (assuming that they are not interested in or aware of privacy issues) and that information privacy is mostly required in their organizational context.

Three interviewees believed that privacy is not important for everyone in the population [10 comments] as "people don't care" if their privacy is breached and people think no one is interested in a "nobody" or a "general person" like them. One interviewee concluded that "for the majority of the people privacy is not an issue". The interviewees also mentioned user awareness issues and associated knowledge asymmetries [9 comments]. They believe that people are not fully aware of privacy implications and issues ("I don't think that companies are not aware of the impact of these systems; it is the individual, sitting in front of it, who is probably not aware of it"), that they "have a very vague notion of what privacy means" and find it difficult and painful to "read and do all the stuff you don't care [about]."

Most of our interviewees observed that information privacy matters more in their working environment.

Only one interviewee said that developers and researchers from his working environment were not interested in privacy concerns [1 comment]: "I have found in my particular role that sometimes it was very difficult to pass the message to the developers or even to the researchers, they were not interested in privacy, or to take those concerns [sic]; you need to have multiple conversation before they are willing to agree to compromise their design decisions to accommodate those privacy features."

In nine remarks that referred to the importance of information privacy as perceived in their organizational context, engineers recognized information privacy as something that is deliberately considered in their respective environment as "there is certainly a lot of thinking about these issues" and people in the companies are "very concerned", "cautious" and "fairly careful" about it [9 comments].

They also referred to information privacy as something that must be dealt with and that is somehow required [11 comments], saying that "it is quite a serious matter", "it has to be there" and that "privacy is not an optional thing anymore". For some, the reasoning behind the consideration of privacy issues is to avoid criticism and a negative public image ("if there is something, if the press was taking [it] down the wrong pipe, then we're dead"; "in general you cannot get very far in collaborations and so on if you don't have that" [this comment refers to "ethics" and "thinking about privacy"]).

In a subsequent survey-based study (Spiekermann, Korunovska, and Langheinrich 2018), we developed

**Control Beliefs**

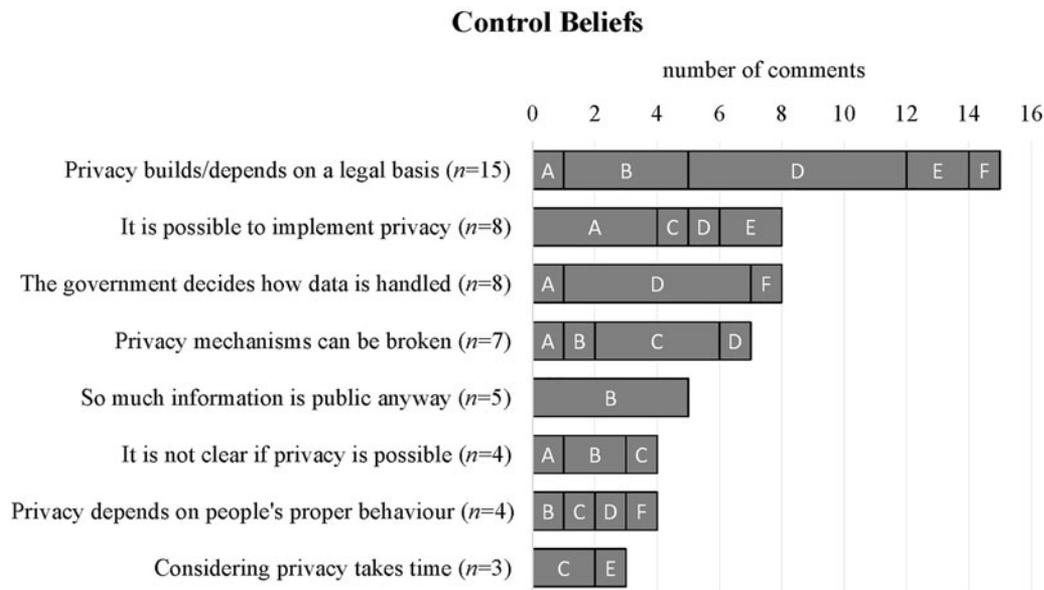

**Figure 6.** The engineers' control beliefs regarding the incorporation of information privacy mechanisms as expressed in 54 comments, ordered by the descending numbers of comments for each belief; the interviewed engineers are anonymously represented within the bars as the letters A to F.

more nuanced insights. Subjective norm was measured with a 5-point differential scale asking engineers whether most people who are important to them think that they should (1) or should not (5) incorporate privacy mechanisms into the systems they build. It turned out that privacy engineering was expected of engineers ($M_{pr} = 4.13$, $SD_{pr} = 1.10$). Only 13 engineers (11%) indicated that the people who they find important would *not* expect them to incorporate privacy mechanisms. That said, we used another item in our questionnaire, which queried engineers' organizational context: we asked about the strength of the normative privacy belief of the engineers' employers (see Appendix 2). Here we got a picture that enriches our qualitative findings while challenging engineers' subjective norm: In fact, only 62% ($n = 77$) of the engineers in our sample work for organizations that expect them to consider privacy mechanisms ($M_{pr} = 3.80$; $SD_{pr} = 1.09$). Thirty eight percent work for employers without clear or even negative privacy norms.

### Control beliefs

In 54 comments, engineers expressed their beliefs with regard to control over privacy implementations (see Figure 6). As with privacy beliefs in general, statements that were categorized as control beliefs were made at various points in the interview, for example in relation to questions about their interpretation of ethical computing or their skills and autonomy. While a few comments indicated that our interviewees believed that it is possible to implement privacy, they also pointed to several difficulties that could reduce their individual control over privacy implementation as engineers. In particular, it turns out that there seems to be a conflict with the legal world with regard to data protection and information privacy.

Five out of six interviewees believed that privacy is a legal issue and that only after the legal issues have been "fixed" – the laws passed and legalities settled – we could talk about the technological implementations [15 comments]. Key statements were as follows: "without a legal framework there is no chance of getting privacy" and "the more liability your corporation has, the more careful it is."

Four interviewees mentioned ways that allow for technological protection of privacy ("there are things that automatically check whether you follow these guidelines; and we also do privacy checks [too] which can be done automatically, for instance if no information should flow out of a program and things like that"; "but we will be able to solve many privacy problems") [8 comments]. However, we can again observe that the same engineers who expressed optimism also express concerns at other points in the interview.

For one thing, they see privacy as entangled with national interests. Three out of six interviewees perceived the government to be an important power that always decides in the end as it has the sovereignty to tell corporations what to do and what data (not) to use ("when corporates are collecting information about their customer base, you are kind of liable to give it to the government at some point, if they ask you to do so"; "as it were, if the government wants my data, then there is a law, that I have to give away my data") [8 comments].

Furthermore, privacy mechanisms can be broken ("there are so many ways of breaking privacy"), overridden ("and every mechanism, that you then build in, can somehow be levered out – and I think most often this will also happen", "it is quite obvious, when you have the right tools and the right data, it doesn't yield you anything"), overruled ("because everyone can easily overrule privacy") and that anonymized data can be de-anonymized with additional information sources ("because everyone knows that maybe with clever tricks you can maybe again deanonymize if I bring in external sources") [7 comments].

One interviewee expressed further concerns by referring to the amount of information that is already available, saying that "everything is quite public" and that "there are so many ways of inferring about the person", which makes the protection of privacy more difficult [5 comments].

Other comments passed the responsibility of control onto the users and their proper behaviour [4 comments], either because they have the choice ("if you don't want to be known you switch off your cell phone") or because they make mistakes ("they don't know the trade-off; and at that point they make mistakes").

Although they expressed the opposite at other points in the interview, some interviewees even doubted the feasibility of privacy per se [4 comments] as "the question of whether privacy is possible or not is still up in the air" and they believe that companies will not easily let go of the data that they could otherwise use or sell.

And lastly, taking privacy into consideration slows down the whole process [3 comments]: "we have to think of the data protection mechanisms and develop them, it certainly would be easier, if we did not have to do that; then we would be faster done with the study and with the whole development of the systems".

The control beliefs we found in our interviews largely point to the larger environment in which privacy is finally achieved or not. Only one external control belief was mentioned that is directly related to the engineers' working environment; that is the time required for building privacy-friendly systems. We tested for this aspect in our quantitative study asking engineers how difficult (1) or easy (5) it would be for them to incorporate privacy mechanisms into their systems in the immediate future (2-3 years). The mean result pointed to time difficulties ($M_{pr} = 2.68$, $SD_{pr} = 1.09$): only 22% of the engineers we asked believe that time is *not* a problem for them when it comes to privacy engineering.

### Perceived behavioural control

The following questions in the interview guide targeted engineers' skills and autonomy: "Could you do more if you really wanted to? Do you have the leeway? Do you have the skill set? Do you have the time?" They shared perceptions of their own behaviour control in 35 comments, which all pointed to a lack of control (see Figure 7). This was due to missing resources and skills as well as (technical) challenges in building privacy-sensitive systems.

Only two interviewees felt that they had the resources, that is, the experience or time to solve privacy issues [2 comments] ("I have worked on privacy"; Question: "Are you considered as a privacy specialist in the organization, so that they give you the time specifically to think about privacy mechanisms?" – Answer: "Yes. I've written papers which discuss privacy, so of course").

All of the interviewees found it difficult to deal with privacy issues and solve them technically ("it is by all means difficult to fulfil certain requirements regarding data storage"; "it's somewhat clumsy and blunt and anything else"; "the design itself is very hard"; "there are several implications in terms of just designing a system that will take privacy and security into concentration which makes it quite hard") [10 comments].

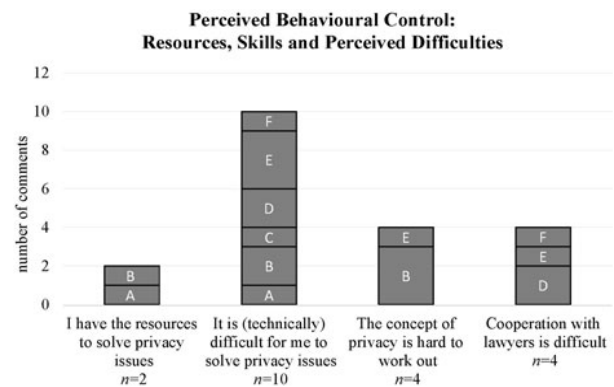

**Figure 7.** The engineers' perceived resources, skills and difficulties in incorporating information privacy mechanisms as expressed in 20 comments; the interviewed engineers are anonymously represented within the bars as the letters A to F.

Furthermore, the concept of privacy is hard to work out [4 comments]: "it is just very hard to figure out when you want information to be revealed and when you do not want it to be revealed", "incredibly hard to define, what is meant by privacy, especially in location", "but there are increasingly some of these softer requirements where there should be humans in the loop to kind of check, those become quite hard to interpret by the developer or the engineer."

What is more, working on privacy often requires cooperation with lawyers, which some of the interviewees found tiresome and difficult [4 comments]: "There are simply people who do not understand the technical realities and make definitions from a legal perspective, that essentially are not reasonable", "I was working with one of the lawyers of our company … it was a nightmare to explain to her certain things and also to know from her the regulations."

A very similar picture emerges with regard to autonomy. Engineers who commented on their autonomy mostly pointed to a lack of autonomy when it comes to decisions on privacy design (see Figure 8).

Two interviewees said that they have the autonomy to solve privacy issues [2 comments] ("the decision was taken by myself"). However, one of them (interviewee "F"), together with interview partner "B", expressed at other points of the interview that they do not have the autonomy to solve privacy issues [8 comments], or had only limited autonomy [5 comments]. They expressed that it "is not up to them" or that they have no final control ("sometimes you get that kind of requests incorporating some of these features, then we have to do it" – in this comment, this engineer also referred to requests that he did "not agree with ethically" such as checking the location or age of users for market research; "you don't really have a choice," "Autonomy exists and double thinking about the implications. But whether you incorporate it into a large scale system, there is no autonomy") and they have only some autonomy ("it's more in the middle", "that is not entirely up to me; there are some other elements too").

Our survey results confirm a control issue among engineers. Thirty seven percent of the systems engineers ($n = 46$) do not feel that they have sufficient control over implementing privacy mechanisms ($M_{pr} = 3.58$, $SD_{pr} = 1.09$). This is *not* due to their capability. Sixty six percent ($n = 82$) said that if they wanted to, they could incorporate privacy mechanisms. Only 26% ($n = 32$) of the engineers believe that they do not have sufficient knowledge to implement privacy. Instead, they face a controllability issue in their work context: Over half of our respondents (51%; $n = 63$) pointed out that in their respective organization it is not (solely) up to them whether they will pursue privacy or not. As outlined above, many seem not to get the time required to implement privacy. But our quantitative study also confirms that autonomy is an issue. Fifty two percent ($n = 64$) say that they do not have the autonomy to implement privacy controls into their systems. Even though the degree of perceived behavioural control over privacy engineering is positively correlated with the hierarchical position: 7% in the higher ranks still express a low level of control (considering their mean perceived behavioural control), and 31% say that with the autonomy they are given it is difficult to implement privacy protection solutions.

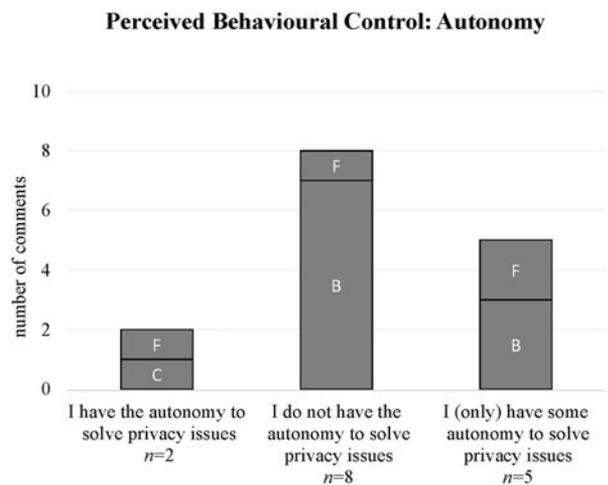

**Figure 8.** The engineers' perceived autonomy to incorporate information privacy mechanisms as expressed in 15 comments; the interviewed engineers are anonymously represented within the bars as the letters A to F.

## Perceived responsibility

Several questions in the interview guide referred to the engineers' perceived responsibility, e.g. "How do you see your own responsibility?" and "What was your role and responsibility in the respective project?". As we can see from the tally of comments shown in Figure 9, the majority of interviewees did not feel responsible.

Only two out of the six interviewees said that they feel responsible for incorporating privacy mechanisms into their systems [7 comments] ("I have the sole responsibility"; "it's a choice I have to make"), but at other points of the interview both of them said that they are not responsible [3 comments] ("but we are not responsible for the product"; "and it is not just me, if I did not develop this system, somebody else

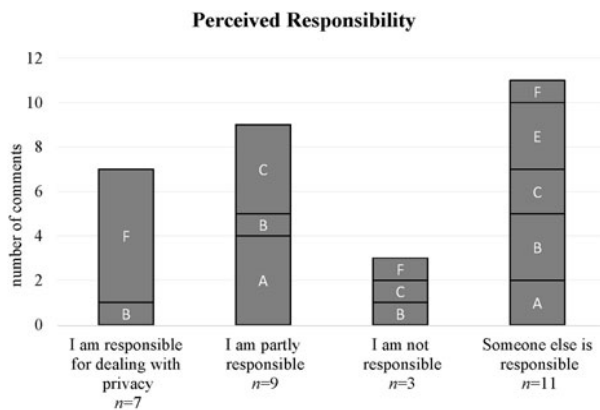

**Figure 9.** The engineers' perceived responsibility as expressed in 30 comments; the interviewed engineers are anonymously represented within the bars as the letters A to F.

will; or at least there are other systems out there which are capable of doing something similar").

Three interviewees felt only partly responsible [9 comments] ("I admittedly have a certain responsibility"; "my part is a really small one in that scale").

Most of their comments pointed to someone else they saw as responsible [11 comments], ranging from the user ("responsibility lies with those that deploy it") to the companies ("it is really up to them"), colleagues ("but I certainly have colleagues; there is for instance a privacy person that works more on the technology side") or the code ("so when we do something like that with companies, we give them the code; so we give them the whole rights for the stuff, so we get rid of everything; then they can do whatever they want").

Our quantitative study points to a similarly nuanced position towards responsibility. Sixty three percent ($n = 77$) of the engineers felt responsible for privacy engineering. We asked whether they agreed that privacy-friendliness is not *their* responsibility. They somewhat disagreed with this ($M = 3.63$; $SD = 1.04$). Notably, engineers in management positions (including the self-managing independent coders) report significantly more responsibility [$F_{pr}$ $(2,114) = 3.10$, $p < .05$]. That said, we would argue that the fact that 37% of the engineers dismissed their responsibility somewhat confirms the mixed views found in the interviews.

## Discussion

We want to emphasize three core findings from our analysis: First, many senior engineers perceive privacy demands as a burden, even though they understand the necessity of taking care of it. Second, they are deeply divided with regard to their control over and responsibility for privacy implementations. Third, they find themselves engaged in an ongoing struggle over information privacy with lawyers.

### Engineers' burden

More than three fourths of all 243 comments on privacy ($n = 188$, 77.4%) were negative, sceptical, or pessimistic – locating the responsibility with other people or listing problems and difficulties associated with the implementation of privacy protections. We found that almost all TPB factors that predict the intention to meet privacy demands (privacy beliefs, experiential attitudes, subjective norm, control beliefs, perceived behavioural control, perceived responsibility) are mostly negative.

The reasons given by our interviewees for their negative beliefs regarding privacy and its implementation are manifold. First, they perceive privacy as a vague concept and its value as uncertain, not always legitimate, context-dependent, and not absolute. It seems that they do not know how to ensure privacy in different contexts in a proper way. Therefore, it is understandable that these beliefs would have a negative effect on their motivation to implement privacy mechanisms. As far back as twenty years ago, the ambiguity on what constitutes privacy was discussed as a "systemic disease" that stymied efforts to protect it (Smith 1994, 167). More recently, the context-dependence of privacy has gained currency in the privacy discourse, drawing Nissenbaum's (2009) work on "contextual integrity" – legitimacy of data use depends heavily on the context of use and is therefore dynamic. However, it is difficult to make sure that systems and data are not used out of context – and engineers know this.

Second, privacy makes things technically more difficult for engineers. The engineers interviewed mentioned resource difficulties in 90% ($n = 18$) of their reflections on past experiences and anticipated obstacles. They said that, on the one hand, consideration of privacy takes a lot of time and, on the other, privacy mechanisms can be broken, overridden or overruled. Furthermore, it is tricky to ensure information privacy as it also depends on the users' behaviour and their vulnerability to getting tricked into revealing personal data. Such shifting of responsibility onto the users had been observed before with app developers, with one developer proclaiming that "at the end of the day its [sic] up to the user" (Greene and Shilton 2017, 14).

Third, despite the senior positions of our interviewees, perceived behavioural control over privacy engineering turned out to be a negative motivational driver. When speaking about their autonomy with regard to design decisions for privacy, 87% ($n = 13$) of their statements indicated that they do not have such autonomy. The reason for this lack of autonomy is not clear from our data. It may be that negative organizational conditions (business models favouring data collection, organizational strategy, time pressure in development, etc.) restrict engineers' degrees of freedom when developing privacy protection mechanisms (Balebako et al. 2014; Berenbach and Broy 2009). Further research into this issue is definitely called for. In sum, our interviewees' responses signal frustration among engineers on matters related to privacy, many of whom even believe that this whole privacy effort is in vain.

### Engineers' inner conflict

Regardless of our interviewees' overall negative emotions and frustrations regarding privacy, they recognize that it is needed and important. Half of the comments in the instrumental attitudes category said something to this effect. But this count is misleading, as most interviewees also contradicted their position at a later point in the interview. They were ambivalent when it came to their perceived behavioural control, which is especially noteworthy as our interviewees were senior engineers who (should) have the knowledge and resources to consider and implement privacy protections in the systems they design.

In one comment each, four out of our six interviewees expressed that they have the resources or the autonomy to solve privacy issues. However, all of them also noted in roughly one third of their statements on control how difficult they find it to implement privacy (ambiguities associated with privacy, technical challenges, and legal complexities). One of the two engineers who mentioned having design autonomy contradicted himself, later saying that he did *not* have the autonomy, or only had *some.* Another engineer mentioned on several occasions that he had neither the choice nor the final control. Such lack of autonomy and control is especially startling as all interviewees hold senior positions and hence should be in the position to strongly influence (if not determine) how privacy is dealt with in their teams and projects.

When it comes to perceived responsibility for privacy, 40% ($n = 12$) of the comments indicated partial responsibility or none at all. Roughly 37% ($n = 11$) of the comments pointed to other responsible parties. Most remarkably, our interviewees again made many self-contradictory remarks, feeling fully or partly responsible for the incorporation of privacy but at the same time mentioning someone else's responsibility or saying "it is not up to me" or something to this effect. Fifteen years ago, Langheinrich and Lahlou (2003) had similar findings, including the comments with similar phrasing. In sum, our findings show that engineers have a deep inner conflict on privacy.

### Engineers' battle with lawyers

At several points in the interviews, engineers mentioned privacy laws as well as the legal staff in their organizations. Our interviewees perceived privacy as a concept that is legally hard to define. Further, they felt that cooperating with lawyers is difficult and tiresome, making it hard to reach a shared level of understanding with them. Most importantly, they were of the belief that the legal basis for privacy has not been settled yet. In their opinion, privacy only made sense once this "legal issue" was fixed and the legal parameters had been clearly established: "without a legal framework there is no chance of getting privacy". Beside the difficulties of communication and collaboration between disciplines, such views of senior engineers of privacy law are alarming, as a fairly well developed framework for privacy regulation has been around since 1980 in the form of the OECD guidelines on the protection of privacy, which was reinforced and expanded in 1995 by the data protection directive 95/46/EC of the European Parliament and the Council of the European Union (1995).

It may be that the EU's GDPR, which recently came into effect in May 2018, will create further clarity for engineers. Interestingly, while engineers pointed at lawyers in our interviews, the same finger-pointing can be observed in the legal world, which is frustrated with engineers' reluctance to embrace privacy. In a recent paper, legal scholars Birnhack, Toch, and Hadar (2014) presented an analysis of computer science educational material and textbooks which continue to promote data collection maximization (instead of privacy-friendly data minimization) and ignore matters of data flow control and privacy.

Taken together, our theoretical and empirical insights suggest that there may be an underlying conflict between the legal world and the engineering world, with lawyers imputing responsibility on engineers that the engineers do not want to embrace. We wonder whether this conflict can be resolved if engineers receive better legal education, learn more about privacy at university, and are better oriented to the long list of hard requirements raining down on them due to new data protection regulations like the GDPR.

## Conclusions

Our findings suggest that engineers deal with privacy related issues, mostly because they are required to do so. On the other hand, all of the senior engineers we interviewed saw difficulties in the implementation of privacy protection measures, which are not only of a technical nature. Moreover, we identified very few clear expressions of responsibility, autonomy, and control in the engineers' statements. Their mostly negative experiential attitude coupled with their awareness of many challenges related to privacy as well as the lack in perceived social pressure from the general population result in an overly negative motivational stance towards Privacy by Design. Where they do not see responsibility for themselves, they see it with the legal world, which they do not like to deal with. These findings are very much in line with the 2003 survey findings reported in Langheinrich and Lahlou (2003). Even though their study is now 15 years old, we still see the same issues with regard to engineers' perceived importance of privacy, the resources available to them, their sense of responsibility, and the autonomy they have when dealing with privacy related issues.

When confronted with a task that is time-intensive, makes things "clumsy" and "very heavy", entails technical difficulties and arduous co-operation with experts from another discipline, engineers have to be driven by a high degree of self-motivation. However, the findings of our interview study point to a low motivation of engineers to deal with privacy related issues. These findings are discouraging, given the rapid rise of personal data markets, data-based discrimination, manipulation, and recurring privacy breaches (Christl and Spiekermann 2016).

If we want to protect human values in an increasingly technological society, we need to find ways to motivate engineers to be sensitive to values such as privacy in their designs. Several approaches have been suggested, such as ethics education for engineers, professional codes of ethics, external ethics experts, and ethical design practices within design teams. Both the educational approach (e.g. Ware, Ahlgren, and Silverman 2013) and incorporation of ethical design practices in laboratories to create "values levers"

(Shilton 2013) seem promising. Professional codes of ethics also have potential to influence engineers' ethical awareness (e.g. Fleischmann, Wallace, and Grimes 2010). However, the strong negative attitude of engineers towards legal experts we observed in our study raise doubt the effectiveness of bringing in external ethical experts.

While the small sample of this interviews-based study demands a cautious interpretation of findings, we see in the responses of our interviewees indicators of likely resistance to what society will increasingly demand beyond technical functionality in the future such as incorporation of privacy mechanisms in the products they design and develop. While the findings of this study are not generalizable, they clearly show that several factors have to be considered as significant influences on the motivation of engineers. So far, studies have focused too narrowly on single factors such as personality and ethics. We hope to encourage more research on dynamics in play in organizations that impede the incorporation of privacy protections in products and services they produce.

## Notes

1. See https://www.acm.org/about-acm/acm-code-of-ethics-and-professional-conduct
2. See https://www.ieee.org/about/ethics.html

## Appendix 1 – Interview Guide

### 1. Introduction

This survey is about the question how we can integrate more consciously ethical decision making into system design and what it is today that makes such integration so difficult. The survey covers your attitudes and experiences as well as organizational issues and time issues; finally your view on engineers' thinking generally (i.e. do engineers see themselves as artists).

### 2. What is "ethical computing" from your perspective?

Some people say it means to build a privacy-sensitive system. But what else would you see as relevant beyond privacy? What constitutes a "good" or "bad" system from a moral perspective? Can you do a quick brainstorming and give a short justification for your ideas?

What is the difference between privacy & security from your perspective?

It would be helpful if you described a system to me where you thought ethical issues were at stake.

What was your role and responsibility in the respective project?

### 3. How do you spontaneously feel about ethical requirements?

Pleasure or nuisance? Rather positive or rather negative?

### 4. User: What would be 3 adjectives or characteristics that you think of when you think of a user of a system you build. Take the system example above. Why do these characteristics come to your mind?

### 5. Attitudes & Beliefs

What are disadvantages and challenges of incorporating privacy mechanisms into your projects? (e.g., code loses its beauty … )

Are security problems more exciting and challenging than privacy problems?

Do you find security problem solving more pleasing and enjoyable?

### 6. Responsibility

Who is responsible for ethical issues in system design?

How do you see your own responsibility?

The responsibility of development teams generally?

Are ethical issue more a matter for legal departments?

Do you have practical ideas on how responsibility could be created?

Do you think that there is room for ethical design decision debates during project development?

In what phases of design do you think such debates could be useful? Please think of each phase: requirements engineering with management, modeling, prototyping, testing.

### 7. Supply chain of software development

What is happening to the systems you build?

What is happening to the code base? Do you share it? Will it be reused?

Could ethical design decision be inherited by those who use your system?

How important do you consider supply chain issues for the responsibility question?

### 8. Organizational Environment [subjective norm]

Most people who are important to me think …

My organization thinks …

Why is this thinking prevalent? How much do you want to comply with what your environment thinks? And why? What is your own thinking?

### 9. Your own skills and autonomy [perceived behavioral control]

Could you do more if you really wanted to?

Do you have the leeway? Do you have the skill set? Do you have the time?

## Appendix 2 – Questionnaire Items

| | |
|---|---|
| **Experiential attitude** | For me the prospect of actually incorporating privacy mechanisms or processes into my new systems in the immediate future (2-3 years) would be … *pleasing - - - - - - - - -annoying* *enjoyable - - - - - - - - - unenjoyably* *exciting - - - - - - - - - boring* *challenging - - - - - - - - - trivial* |
| **Instrumental attitude** | I find that incorporating privacy mechanisms into the design of my systems in the immediate future (2-3 years) *up-to-date - - - - - - - - - outmoded* *very useful - - - - - - - - - useless* *sensible - - - - - - - - - senseless* *fruitful - - - - - - - - - futile* *valuable - - - - - - - - - worthless* |
| **Subjective Norm** | Most **people who are important to me** think that *I should - - - - - I should not* incorporate privacy mechanisms into the systems I build |
| **Normative Beliefs of the Organization** | Against the background of your respective organizational context (company, university, research group), what is true for you? My **organization** thinks that *I should - - - - - I should not* incorporate privacy mechanisms into the systems I build |
| **Perceived Behavioural Control** | It is mostly up to me whether or not I incorporate privacy mechanisms into the systems I build in the immediate future (2-3 years). *strongly agree - - - - - - - - - strongly disagree* If I wanted to I could incorporate privacy mechanisms into the systems I build in the immediate future (2-3 years). *definitely true - - - - - - - - - definitely false* |
| **Control Beliefs** | **The knowledge** I need to have to incorporate privacy mechanisms into my systems would make it *very difficult - - - - - very easy* for me to do so in the immediate future (2-3 years). **The time** required to incorporate privacy mechanisms into my systems would make it *very difficult - - - - - very easy for me* to do so in the immediate future (2-3 years). **The autonomy** I need to have to incorporate privacy mechanisms into my systems would make it *very difficult - - - - - very easy* for me to do so in the immediate future (2-3 years). |
| **Responsibility** | Ensuring the privacy-friendliness of a system is not my responsibility. *strongly agree - - - - - - - - -strongly disagree* |

## Appendix 3 – Engineers' Beliefs and Correlations with Privacy Attitudes

| Political (PB) and Technical Beliefs (TB) about Privacy | M | SD | Instrumental attitudes | Experiential attitudes |
|---|---|---|---|---|
| 1. PB: Designing user-privacy systems into systems is important to enable a power balance between CORPORATIONS and citizens | 4.12 | 0.98 | 0.28** | 0.22* |
| 2. PB: Designing user-privacy into systems is important to enable a power balance between GOVERNMENTS and citizens | 3.94 | 1.02 | 0.16 | 0.05 |
| 3. PB: I think that more data means more knowledge | 3.60 | 1.10 | 0.06 | 0.02 |
| 4. PB: I think that personal information has become just another form of property that people can sell or buy | 3.41 | 1.33 | 0.03 | 0.00 |
| 5. PB: I think that freedom of speech is more important than privacy | 3.09 | 1.07 | −0.22 | −0.15 |
| 6. PB: I think that transparency is more important than privacy | 3.00 | 1.11 | −0.36* | −0.41** |
| 7. TB: Ensuring user-privacy in a system is a legal issue rather than a technical one | 2.95 | 1.27 | −0.12 | 0.07 |
| 8. TB: I think that technology is neutral | 2.88 | 1.43 | −0.05 | −0.08 |
| 9. TB: Efforts to fully secure a system are often futile, because good hackers can circumvent any security | 2.81 | 1.31 | −0.10 | −0.04 |
| 10. TB: I think that with the right cryptographic mechanisms most privacy problems can be solved | 2.44 | 1.24 | −0.01 | −0.07 |
| 11. TB: As Ubiquitous Computing systems inherently rely on the collection of large amounts of data, privacy and UbiComp is a contradiction | 2.43 | 1.11 | −0.13 | −0.12 |
| 12. TB: I think that it is possible, in principle, to build error-free systems | 2.21 | 1.28 | −0.06 | 0.06 |